\documentclass[aps,pre,twocolumn,showpacs,superscriptaddress]{revtex4-1}

\usepackage{amsmath}
\usepackage{amssymb}
\usepackage{graphicx}

\begin{document}

\title{Quantification of  noise in the bifunctionality-induced post-translational modification}

\author{Alok Kumar Maity}
\affiliation{Department of Chemistry,
University of Calcutta, 92 A P C Road, Kolkata 700 009, India}

\author{Arnab Bandyopadhyay}
\affiliation{Department of Chemistry,
Bose Institute, 93/1 A P C Road, Kolkata 700 009, India}

\author{Sudip Chattopadhyay}
\email{sudip$_$chattopadhyay@rediffmail.com}
\affiliation{Department of Chemistry,
Bengal Engineering and Science University, Shibpur, Howrah 711103, India}

\author{Jyotipratim Ray Chaudhuri}
\email{jprc$_$8@yahoo.com}
\affiliation{Department of Physics,
Katwa College, Katwa, Burdwan 713130, India}

\author{Ralf Metzler}
\email{rmetzler@uni-potsdam.de}
\affiliation{Institute for Physics \& Astronomy, University of Potsdam,
D-14476 Potsdam-Golm, Germany}
\affiliation{Physics Department, Tampere University of Technology,
FI-33101 Tampere, Finland}

\author{Pinaki Chaudhury}
\email{pinakc@rediffmail.com}
\affiliation{Department of Chemistry,
University of Calcutta, 92 A P C Road, Kolkata 700 009, India}

\author{Suman K Banik}
\email{Corresponding author; skbanik@jcbose.ac.in}
\affiliation{Department of Chemistry,
Bose Institute, 93/1 A P C Road, Kolkata 700 009, India}

\date{\today}

\begin{abstract}
We present a generic analytical scheme for the quantification of fluctuations due to
bifunctionality-induced signal transduction within the members of bacterial two-component
system. The proposed model takes into account post-translational modifications in terms of
elementary phosphotransfer kinetics. Sources of fluctuations due to autophosphorylation,
kinase and phosphatase activity of the sensor kinase have been considered in the model
via Langevin equations, which are then solved within the framework of linear noise
approximation. The resultant analytical expression of phosphorylated response regulators
are then used to quantify the noise profile of  biologically motivated single and branched
pathways. Enhancement and reduction of noise in terms of extra phosphate outflux and
influx, respectively, have been analyzed for the branched system. Furthermore, role of
fluctuations of the network output in the regulation of a promoter with random
activation/deactivation dynamics has been analyzed.
\end{abstract}

\pacs{87.18.Mp, 87.18.Tt, 87.18.Vf}

\maketitle


\section{Introduction}

The response of living systems to an external stimulus is coordinated by
highly specialized signal transduction machinery. In the bacterial kingdom, this is achieved
by the well characterized two-component system (TCS) minimally comprised of the membrane
bound sensor kinase (SK) and the cytoplasmic response regulator (RR)
\cite{Appleby1996,Hoch2000,Laub2007,Hart2013}. The machinery of TCS is utilized
by the bacteria to
process the information of external signal in terms of phosphotransfer kinetics.
When applied, an external stimulus causes phosphorylation at the histidine residue
of SK which then gets transferred to the cognate (and/or non-cognate) RR at its aspartate
domain. The phosphorylated RR then acts as a transcription factor for several downstream
genes, as well as for the activation/represion of its own operon. It is now a well established fact
that in addition to being a source (kinase), some SK can also act as a sink (phosphatase) while
interacting with an RR \cite{Hsing1998,Laub2007,Goldberg2010,Hart2013}. Such bifunctional
behavior of SK towards RR can altogether build a robust motif in the bacterial signal transduction
network \cite{West2001,Batchelor2003,Shinar2007,Siryaporn2010}.

The expression of proteins in individual cells is usually driven by the fluctuations present
within the cellular environment as well as the fluctuations imposed by the external stimulus
\cite{Eldar2010,Elowitz2002,Munsky2012,Paulsson2004,Rosenfeld2005,SilvaRocha2010,
Thattai2001}. This often leads to variability in the expression level within the context of a
single cell \cite{Davidson2008,Losick2008,Rotem2010,Sureka2008}. When observed in
the bulk, such fluctuations get averaged out over the cellular population.
The prevalent fluctuations, whether external or internal, not only effect the dynamics
of gene expression, but also play a major role in post-translational modification
\cite{Jia2010,Mehta2008}. In this connection, it is also important to mention
the role of cellular fluctuations in the different signal transduction motif that primarily uses
phosphotransfer mechanism. Using a push-pull amplifier loop mechanism, theoretical study
has been made to analyze the signal transduction within the photoreceptor of retina
\cite{Detwiler2000}. Theoretical model has been proposed to study the effect of reversibility
in the phosphorylation-dephosphorylation cycle that can generate bistable behavior in the
presence of noise and can propagate within the signaling cascade \cite{Miller2008}. In the
context of robustness in the bacterial chemotaxis, reversibility on a signaling cascade has
been shown to exert a stabilizing effect of adaptation through methylation \cite{Alon1999}.
Correlation between extrinsic and intrinsic noise due to external signal and internal
biochemical pathways, respectively, has also been reported to enhance the robustness
of zero-order ultrasensitivity \cite{TanaseNicola2006}.

Post-translational modification in terms of phosphate transfer is important to generate
the pool of phosphorylated RR that acts as a transcription factor
for several downstream genes. Bifunctionality, on the other hand, plays a crucial role
to maintain this pool as the information flows through the phosphotransfer motif. Thus
bifunctionality and post-translational modification work hand in hand to maintain the
optimal pool of phosphorylated RR. Since this composite functional behavior takes place
in a noisy cellular environment, it is worthwhile to investigate the role of cellular noise on the
bifunctionality controlled post-translational modification of the components of the well
composed TCS signal transduction machinery. The above observations have motivated us
to develop a general model to quantify the molecular noise in the bacterial TCS considering
both the bifunctional SK and the post-translational modification of RR. The proposed model
takes care of the elementary stochastic phosphotransfer kinetics between the two members
(SK and RR) of the TCS and gives a prescription to calculate the fluctuations associated with
the phosphorylated RR keeping in mind the bifunctional property of the TCS. We further
investigate the role of fluctuations of the network output in the regulation of a promoter
with random activation/deactivation kinetics.


\begin{figure}[!t]
\begin{center}
\includegraphics[width=0.75\columnwidth,angle=0]{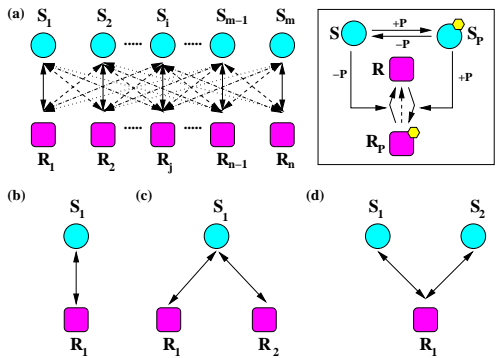}
\end{center}
\caption{(color online) Wiring diagrams for the proposed post-translational
interactions between SK and RR. The cyan circle and magenta square stand 
for SK and RR, respectively.
(a) Generalized $m$:$n$ system with cognate (solid arrow) and
non-cognate (dotted, dashed and dotted-dashed arrow) kinase and phosphatase interactions.
(b), (c) and (d)  are wiring diagrams for 1:1, 1:2 and 2:1 interactions, respectively.
The boxed wiring diagram depicts the phosphate addition (+P) and removal
(-P) kinetics between a pair of cognate and/or non-cognate SK and RR that ultimately
results into phosphorylated RR (magenta square with a yellow hexagon on top). 
The dotted arrowhead is for auto-dephosphorylation of RR. For simplicity, we do not
show the yellow hexagon in (a)-(d).
}
\label{general}
\end{figure}


\section{The Model}

We start by considering a simple system describing post-translational modification
driven by phosphotransfer mechanism of a typical TCS, where $m$ numbers of SK
interact with $n$ numbers of RR, the ultimate product of which is $R_p$, the phosphorylated
RR. We call the proposed model as the $m$:$n$ system (Fig.~\ref{general}a) where each
of the SK, RR and their phosphorylated forms are designated as $S$, $R$, $S_p$ and
$R_p$, respectively. The generic model considered here involves single pair interaction
(Fig.~\ref{general}b). In addition, it takes care of \textit{branched pathways}
\cite{Laub2007}; for example the 1:2 system (Fig.~\ref{general}c) mimics the
\textit{one-to-many} pathway as observed in chemotaxis system in \textit{E. coli}, where
the SK CheA phosphorylates two RR, CheY and CheB \cite{Armitage1999}. Similarly, the
2:1 system (Fig.~\ref{general}d) follows the kinetics of \textit{many-to-one} pathway
as observed in \textit{V. cholerae} where the SK LuxS and CqsS phosphorylate
the RR LuxO \cite{Waters2005}.

As mentioned earlier, in typical bacterial TCS, the key steps of phosphotransfer mechanism
involve autophosphorylation at SK, transfer of phosphate group from SK to RR, and
SK mediated removal of phosphate group from RR (see boxed diagram in
Fig.~\ref{general}). To keep the model simple, we do not consider the synthesis (birth) or
degradation (death) of any system component. The interaction we consider here may be
of cognate and/or non-cognate type. For $m$:$n$ pair, one can consider the specific
interaction between $i$-th SK and $j$-th RR, where $1\leqslant i \leqslant m$ and $1\leqslant j \leqslant n$, to write down the elementary kinetic steps considering the minimal interaction
between a specific pair
\begin{subequations}
\begin{eqnarray}
S_i & \overset{\alpha_{i}}{\underset{\beta_i}{\rightleftharpoons}} & S_{pi} ,
\label{eq1a} \\
S_{pi} + R_j & \overset{\gamma_{ij}}{\longrightarrow} & S_i + R_{pj} ,
\label{eq1b} \\
S_i + R_{pj} & \overset{\mu_{ij}}{\longrightarrow} & S_i + R_j .
\label{eq1c} \\
R_{pj} & \overset{\nu_j}{\longrightarrow} & R_j .
\label{eq1d}
\end{eqnarray}
\end{subequations}

\noindent  In the above kinetic steps, Eq.~(\ref{eq1a}) considers autophosphorylation
at the histidine residue of the SK. Generally, autophosphorylation takes place under the
influence of an external signal \cite{Appleby1996,Hoch2000,Laub2007} which we consider
to be of constant type
and is absorbed in the rate constant $\alpha_i$. Eqs.~(\ref{eq1b}-\ref{eq1c}) take
into account the kinase and phosphatase activity of the SK, respectively, thus considering
the bifunctional behavior of the SK. Note that in Eq.~(\ref{eq1c}), the SK acts as an enzyme
to control the dephosphorylation of RR, hence itself remains unchanged
\cite{Hoch2000,West2001,Laub2007}. Eq.~(\ref{eq1d}) denotes the auto-dephosphorylation
of the RR independent of the phosphatase effect of SK on RR \cite{Siryaporn2010}.

Due to the inherent noisy nature of the cellular environment, each of the four reactions
mentioned above are influenced by fluctuations and in turn affect the copy numbers of
each system component. To take this into account, we introduce Langevin noise terms
that can influence each of the reactions independently given by Eqs.~(\ref{eq1a}-\ref{eq1d}).
The interaction of a single SK with multiple RR or \textit{vice versa} in presence of
fluctuations considered here can be compared with stochastic system-reservoir formalism
where a single system interacts with multiple reservoir or \textit{vice versa} \cite{Popov2007}.
The stochastic differential equations describing
the phosphorylated SK and RR in presence of fluctuations can be written as
\begin{subequations}
\begin{eqnarray}
\frac{d S_{pi}}{dt} & = & \alpha_i ( S_{Ti} -  S_{pi} ) - \beta_i S_{pi}
- \sum_{j=1}^n \gamma_{ij} S_{pi} \nonumber \\
&& \times ( R_{Tj} - R_{pj} ) + \xi_{S_{pi}} (t)
\label{eq2a} \\
\frac{d R_{pj}}{dt} & = & \sum_{i=1}^m  \gamma_{ij} S_{pi} ( R_{Tj} - R_{pj} )
- \sum_{i=1}^m [ \mu_{ij} \nonumber \\
&& \times ( S_{Ti} -  S_{Pi} ) + \nu_j  ] R_{pj} + \xi_{R_{pj}} (t) .
\label{eq2b}
\end{eqnarray}
\end{subequations}

\noindent
Here, $S_{Ti} =S_i+S_{pi}$ and $R_{Tj} =R_j+R_{pj}$ stand for
the total amount of $i$-th SK and $j$-th RR, respectively.
The additive noise terms $\xi_{S_{pi}}$ and $\xi_{R_{pj}}$ take care of the fluctuations
in the copy number of $S_{pi}$ and $R_{pj}$, respectively.
Within the framework of linear noise approximation, we define the statistical properties
of the Langevin terms obeying the fluctuation-dissipation relation
\cite{Detwiler2000,Elf2003,Swain2004,Bialek2005,Mehta2008,Kampen2005}
with zero mean, $\langle \xi_{S_{pi}} (t) \rangle =  \langle \xi_{R_{pj}} (t) \rangle = 0$ and
\begin{eqnarray*}
\langle \xi_{S_{pi}} (t) \xi_{S_{pi}} (t+\tau) \rangle & = & 2 \alpha_i ( S_{Ti} -  \langle S_{pi} \rangle)
\delta (\tau) , \\
\langle \xi_{R_{pj}} (t) \xi_{R_{pj}} (t+\tau) \rangle & = & 2 \sum_{i=1}^m  \gamma_{ij}
\langle S_{pi} \rangle ( R_{Tj} - \langle R_{pj} \rangle )
\delta (\tau) ,
\end{eqnarray*}

\noindent
with $\langle S_{pi} \rangle$ and $\langle R_{pj} \rangle$ being the mean values at the
steady state. In addition, the noise terms are correlated \cite{TanaseNicola2006,Swain2004}
\begin{eqnarray*}
\langle \xi_{S_{pi}} (t) \xi_{R_{pj}} (t+\tau) \rangle & = &  -  \gamma_{ij} \langle S_{pi} \rangle
( R_{Tj} - \langle R_{pj} \rangle ) \delta (\tau)  .
\end{eqnarray*}

\noindent
Since the stochastic Langevin equations (\ref{eq2a}-\ref{eq2b}) are nonlinear in nature,
it is difficult to solve them analytically. To make the solution tractable analytically, we employ
linearization of the stochastic equations.
The Langevin equation with the linear noise approximation is a valid approach provided
the input signal is very small. In addition, such linearization also remains also valid when the time
to reach the steady state is longer than the characteristic time scale of the birth and death rate
of the system components\cite{Paulsson2004,TanaseNicola2006,Hu2011}.
Thus, linearizing Eqs.~(\ref{eq2a}-\ref{eq2b}) around the steady state, i.e.,
$S_{pi}=\langle S_{pi} \rangle+\delta S_{pi}$ and
$R_{pj}=\langle R_{pj} \rangle+\delta R_{pj}$, we have
\begin{eqnarray}
\frac{d}{dt}
\left(
\begin{array}{c}
\delta S_{pi} \\
\delta R_{pj}
\end{array}
\right)
& = &
\left(
\begin{array}{cc}
-a_{i} & \sum_{j=1}^n \gamma_{ij} \langle S_{pi} \rangle \\
\sum_{i=1}^m b_{ij}  & -  c_j
\end{array}
\right) \nonumber \\
&& \times
\left(
\begin{array}{c}
\delta S_{pi} \\
\delta R_{pj}
\end{array}
\right)
 +
\left(
\begin{array}{c}
\xi_{S_{pi}} \\
\xi_{R_{pj}}
\end{array}
\right) ,
\label{eq3}
\end{eqnarray}

\noindent where
\begin{eqnarray*}
a_{i}  & = & \frac{
\alpha_i S_{Ti}
}{
\langle S_{pi} \rangle
} ,
b_{ij} = ( \mu_{ij} S_{Ti} + \nu_j ) \frac{
\langle R_{pj} \rangle
}{
\langle S_{pi} \rangle
}, \\
c_j & = & \sum_{i=1}^m \frac{
[ \mu_{ij} ( S_{Ti} - \langle S_{pi} \rangle ) + \nu_j ] R_{Tj}
}{
R_{Tj} - \langle R_{pj} \rangle
} .
\end{eqnarray*}

\noindent Solving Eq.~(\ref{eq3}) and performing Fourier transformation
$\delta \tilde{X} ( \omega ) =  {\int_{- \infty}^{\infty}} \delta X( t ) e^{-i \omega t} dt$
on the resultant solution, we have in matrix notation, the generalized solution for
both $\delta \tilde{S}_p (\omega)$ and $\delta \tilde{R}_p (\omega)$,
\begin{subequations}
\begin{eqnarray}
\boldsymbol{\delta} {\mathbf{\tilde{S}_{p} } ( \omega )}
& = & \mathbf{A}^{-1}
\left (
\mathbf {\langle S_{pK} \rangle}
\boldsymbol{\delta} \mathbf{\tilde{R}_{p}} ( \omega )
+
\boldsymbol{\tilde{\xi}_{S_p}} ( \omega )
\right ) ,
\label{eq4a}\\
\boldsymbol{\delta} \mathbf{\tilde{R}_{p} ( \omega )}\label{eq10}
& = & \mathbf{P^{- 1}} \left (
\mathbf{B}^{\mathrm{T}} \mathbf{A}^{-1}
\boldsymbol{\tilde{\xi}} (\omega) +  \boldsymbol{\tilde{\xi}_{R_p}} (\omega)
\right ) ,
\label{eq4b}
\end{eqnarray}
\end{subequations}

\noindent where
$\mathbf{P} =
\mathbf{C} - \mathbf{B}^{\mathrm{T}} \mathbf{A}^{-1} \mathbf{\langle S_{pK} \rangle}$.
In the above expressions (\ref{eq4a}-\ref{eq4b}),
$\boldsymbol{\delta} \mathbf{\tilde{S}_{p}}$ and
$\boldsymbol{\delta} \mathbf{\tilde{R}_{p}}$ are $m\times1$ and $n\times1$ dimensional column
vectors with elements ${\delta}{\tilde{S}_{pi}}$ and ${\delta}{\tilde{R}_{pj}}$, respectively.
Likewise, $\boldsymbol{\tilde{\xi}_{S_p}}$ and
$\boldsymbol{\tilde{\xi}_{R_p}}$ are $m\times1$ and $n\times1$ column vectors with elements
$\tilde{\xi}_{S_{pi}}$ and $\tilde{\xi}_{R_{pj}}$, respectively.
${\mathbf{A}}$ and ${\mathbf{C}}$ are diagonal matrix of order $m \times m$ and
$n \times n$ with elements
$(i \omega + a_{i})$ and $(i \omega + c_j)$, respectively.
Additionally, $\mathbf{\langle S_{pK} \rangle}$ and $\mathbf{B}$ are matrix of order
$m\times n$ with
elements $\langle S_{pi} \rangle \gamma_{ij}$ and $b_{ij}$, respectively.


\section{Results and Discussion}

\begin{figure}[!t]
\begin{center}
\includegraphics[width=0.75\columnwidth,angle=-90,bb=48 62 590 792]{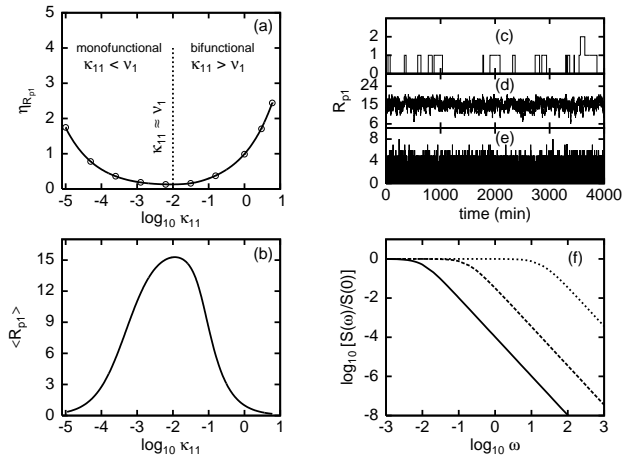}
\end{center}
\caption{Plot of noise, steady state protein level, time series and power
spectra for 1:1 system.
(a) Noise $\eta_{R_{p1}}$ as a function of $\log_{10} \kappa_{11}$.
The noise profile has been shown to contain three different regions,
\textit{viz} monofunctional region ($\kappa_{11} < \nu_1$), crossover region
($\kappa_{11} \approx \nu_1$) and bifunctional region ($\kappa_{11} > \nu_1$).
In the absence of auto-dephosphorylation kinetics (Eq.~(\ref{eq1d})), only the bifunctional
domain exists, whereas in the absence of phosphatase kinetics (Eq.~(\ref{eq1c})), only the
monofunctional domain becomes prevalent. The open circles are due to Gillespie
simulation \cite{Gillespie1976,Gillespie1977}.
For comparison of the noise profile with the Fano factor, see Fig.~\ref{fano11}.
(b) Steady state $R_{p1}$ as a function of $\log_{10} \kappa_{11}$.
(c)-(e) Time series of $R_{p1}$ for low ($\kappa_{11}=10^{-5}$), intermediate
($\kappa_{11}=10^{-2}$) and high ($\kappa_{11}=1$) values of $\kappa_{11}$,
respectively, generated using Gillespie algorithm \cite{Gillespie1976,Gillespie1977}.
Note that in (d) the ordinate does not start from zero.
(f) Normalized power spectra for low (solid line), intermediate (dashed line)
and high (dotted line) values of $\kappa_{11}$.
In all the cases, $\alpha_1/\beta_1=5$, $\nu_1=0.01$ and $S_{T1}=R_{T1}=20$.
}
\label{noise11}
\end{figure}

\begin{figure}[!t]
\begin{center}
\includegraphics[width=0.75\columnwidth,angle=-90]{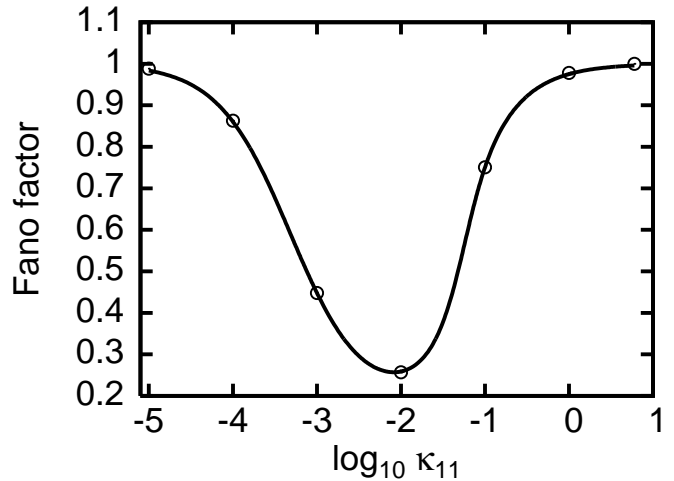}
\end{center}
\caption{Plot of Fano factor as a function of $\log_{10} \kappa_{11}$ for 1:1 system.
The solid line is due to the theoretical calculation and the open circles are generated
using Gillespie simulation \cite{Gillespie1976,Gillespie1977}.
The values of other parameters are same as in Fig.~\ref{noise11}.
}
\label{fano11}
\end{figure}

Since we are interested in the effect of noise on phosphorylated RR, $R_p$, which acts
as transcription factor for one or more genes including the gene that encodes SK and RR, we
now focus on the solution of Eq.~(\ref{eq4b}) only. From the structure of Eq.~(\ref{eq4b}), it
is clear that the dynamics of $R_p$ is now decoupled from $S_p$. To understand the role of fluctuations in phosphotransfer processes,
we define noise at steady state,
\begin{eqnarray*}
\eta_{R_{P} } = \sigma_{R_p} / \langle R_p \rangle,
\end{eqnarray*}
\noindent
where $\sigma_{R_p}$ is the standard deviation of $R_p$ (see Fig.~\ref{noise11}).
It is important to mention that at times fluctuations in the biological systems are quantified 
using Fano factor, $\sigma_{R_p}^2 / \langle R_p \rangle$, where $\sigma_{R_p}^2$ is 
the variance of $R_p$ (see Fig.~\ref{fano11}). Note that in the rest of the discussion 
we have analyzed our results in terms of steady state noise $\eta_{R_{P} }$ only.

While calculating the noise for the three different systems (1:1, 1:2 and 2:1) mentioned
in Fig.~\ref{general}, we only focus on the noise level of $R_{p1}$, the phosphorylated form
of $R_1$. Noise generated due to other interactions ($S_1$ and $R_2$, and $S_2$ and $R_1$)
are considered to add extra layers of information on the noise profile of $R_{p1}$.
During the calculation of noise and power spectra we have considered
$\gamma_{ij} \approx \mu_{ij} \approx \kappa_{ij}$
$(i,j=1,2)$ for simplicity in the strong limit of protein-protein interaction between SK and RR
\cite{Siryaporn2010}.
It is important to note that while interacting with its partner, an SK
shows both monofunctional and bifunctional behavior for $\nu_{j} > \kappa_{ij}$ and
$\nu_j < \kappa_{ij}$, respectively. At $\nu_j \approx \kappa_{ij}$
(cross over regime), the system makes a transition from mono- to bifunctional domain.


In Fig.~\ref{noise11}a noise profile of $R_{p1}$ has been shown in a semilog plot. For 1:1
system, at a low value of $\kappa_{11}$, noise has a nonzero value which goes down
as $\kappa_{11}$ value increases. As $\kappa_{11}$ value increases further, noise
increases and reaches a high value. As evident from the definition, noise is inversely
proportional to the population
of steady state $R_{p1}$. To check the role of $\langle R_{p1} \rangle$ on steady state noise, we
have calculated $\langle R_{p1} \rangle$ as a function of  $\log_{10} \kappa_{11}$
(Fig.~\ref{noise11}b) from which it is evident that the protein profile develops exactly in a way opposite
to the noise profile and imparts an inverse effect on the noise development. For a low
value of $\kappa_{11}$, the auto-dephosphorylation $\nu_1$ dominates over the phosphatase activity of SK on RR
($\kappa_{11} < \nu_1$).  In this regime, the sensor shows monofunctional activity by acting as a
kinase only, which is still lower than $\nu_1$. This effectively reduces the $R_{p1}$
level (Fig.~\ref{noise11}c) and increases the noise of the system.
In the limit $\kappa_{11} \approx \nu_1$ (vertical dotted line in Fig.~\ref{noise11}a), $R_{p1}$ level
reaches its maximum (Fig.~\ref{noise11}b and Fig.~\ref{noise11}d) while reducing the noise.
When $\kappa_{11}$ exceeds $\nu_1$ ($\kappa_{11} > \nu_1$), the phosphatase activity of SK starts
to show up in addition to its kinase activity. In this regime, the bifunctional property of SK comes into
play reducing the copy of $R_{p1}$ (Fig.~\ref{noise11}e) henceforth increasing the
noise of the system.
To compare the noise profile of $R_{p1}$ given in Fig.~\ref{noise11}a with the Fano factor
($\sigma_{R_{p1}}^2 / \langle R_{p1} \rangle$) of the same quantity, we have shown the
dependence of Fano factor on $\kappa_{11}$ in Fig.~\ref{fano11}.
To understand how the system relaxes under the influence of the protein-protein interaction,
we calculate the power spectra
$S( \omega ) = \langle
\delta \tilde{R}_{p1} (\omega ) \delta \tilde{R}_{p1} (\omega' ) \rangle$
(Fig.~\ref{noise11}f). The resultant spectral lines are plotted for low, intermediate and high
$\kappa_{11}$ values. As expected, the power spectra relaxes faster for a low $\kappa_{11}$
value compared to an intermediate $\kappa_{11}$ value which again relaxes faster compared to
a high $\kappa_{11}$ value. For a low value of $\kappa_{11}$, the conversion of $R$ into
$R_p$ is a slow process and hence fast fluctuations in the copy number have minimal effect
on the power spectrum. As $\kappa_{11}$ value increases, the conversion rate increases and
thus gets affected by the noise in the copy number which results into slower relaxation.


\begin{figure}[!t]
\begin{center}
\includegraphics[width=0.75\columnwidth,angle=-90,bb=48 62 590 792]{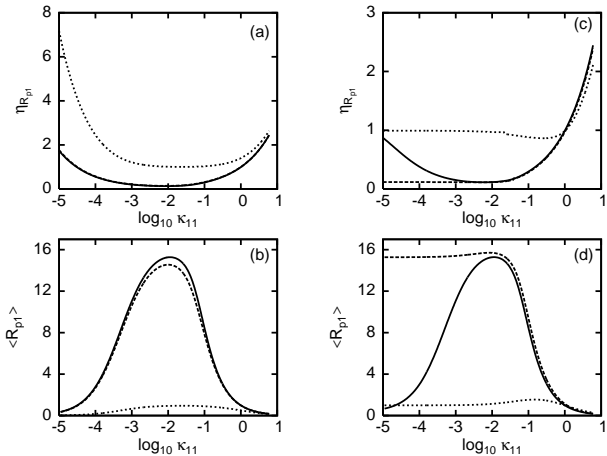}
\end{center}
\caption{Plot of noise and steady state protein level for branched (1:2 and 2:1)
pathways.
(a,c) Noise profile for 1:2 and 2:1 system, respectively,  as a function of $\log_{10} \kappa_{11}$.
The solid, dashed and dotted lines are for low ($10^{-5}$), intermediate ($10^{-2}$) and high ($1$)
values of $\kappa_{12}$ (for 1:2) and $\kappa_{21}$ (for 2:1). (b,d) Steady state level of $R_{p1}$
for the same values of $\kappa_{12}$ and $\kappa_{21}$ as in (a,c).
For all the cases, $\alpha_i/\beta_i=5$, $\nu_j=0.01$ and $S_{Ti}=R_{Tj}=20$.
}
\label{noisebp}
\end{figure}

In Figs.~\ref{noisebp}a-\ref{noisebp}c, we show the noise,
$\eta_{R_{p1}}$ for 1:2 and 2:1 system as a function of $\kappa_{11}$ in a semilog plot. 
For comparison, we refer to noise profile of 1:1 system shown in Fig.~\ref{noise11}a.
In 1:2 system, a single SK, $S_1$ interacts with two RRs, $R_1$ and $R_2$, with its bifunctional
properties acting on both the RRs. In Fig.~\ref{noisebp}a, we show the noise generated for
$R_{p1}$ while considering the kinase and phosphatase rates
($\gamma_{12} \approx \mu_{12} \approx \kappa_{12}$) between $S_1$ and $R_2$ to be low
($\kappa_{12}=10^{-5}$), intermediate ($\kappa_{12}=10^{-2}$) and high ($\kappa_{12}=1$).
For low and intermediate $\kappa_{12}$ values, the noise profile looks almost
that of like 1:1 system as $\kappa_{11}$ is varied. This happens as the interaction between $S_1$ and $R_2$
adds a weak layer of information on $R_1$ due to monofunctional property of $S_1$ on $R_2$
($\nu_2 \geqslant \kappa_{12}$). On the other hand, for a high $\kappa_{12}$ value, a huge amplification of noise
occurs (the dotted line in Fig.~\ref{noisebp}a). In this domain, as $\nu_2 < \kappa_{12}$, SK
starts to show its bifunctional property and is more active in its interaction with $R_2$,
rather than with $R_1$. Such active interaction between $S_1$ and $R_2$ adds an extra
layer of outflux of phosphate group from $R_1$ (the dotted line in Fig.~\ref{noisebp}b) thus
leading to a low level of $\langle R_{p1} \rangle$ and an enhancement of noise.


In 2:1 system, a single RR, $R_1$ interacts with two SKs, $S_1$ and $S_2$. In
Fig.~\ref{noisebp}c, we show the noise generated for $R_{p1}$ while taking into account the
kinase and phosphatase rates ($\gamma_{21} \approx \mu_{21} \approx \kappa_{21}$)
between $S_2$ and $R_1$ to be low ($\kappa_{21}=10^{-5}$), intermediate
($\kappa_{21}=10^{-2}$) and high ($\kappa_{21}=1$).
For low value of $\kappa_{21}$, the noise profile is similar to that of a 
1:1 system as $\kappa_{11}$ is increased. Although in this domain $S_2$ acts as kinase
only, it provides a low level of input on $R_1$ as $\nu_1 > \kappa_{21}$. Interesting
behavior emerges as $\kappa_{21}$ takes an intermediate and high values.
In the intermediate domain, maximal level of $\langle R_{p1} \rangle$ is produced
due to extra influx of phosphate group. This large influx due to $\kappa_{21}$ can overpower
the low kinase effect of $\kappa_{11}$, and hence increase the steady
state level of $R_{p1}$ as a effect of which the noise level attains a minimum value.
For high $\kappa_{21}$, $\nu_1 < \kappa_{21}$ where $S_2$ starts to show its bifunctional property
via phosphate input and removal. This helps the composite system to maintain a high value
of noise for a wide range. Note that, compared to the low and intermediate domain, the protein
level in this region goes down drastically due to strong phosphatase activity of $S_2$ on
$R_1$. It is interesting to note that for intermediate and high $\kappa_{21}$ value, the composite
system loses it monofunctional behavior almost completely (Fig.~\ref{noisebp}d).

To check the effect of the network in the regulation of the downstream promoter activity,
one needs to consider the fluctuations in $R_p$ level as extrinsic noise while the
promoter activation/inactivation is governed by the intrinsic molecular fluctuations
\cite{Hu2011,Hu2012}. The time scale for the relaxation of the network is given by 
$\tau_{in}$ where $\tau_{in} = c_j^{-1}$. The promoter switching kinetics, driven by 
output of the network (i.e., $R_p$), can be modeled as
\begin{equation}
\label{eq5}
Off  \overset{k_{on} R_p (t)}{\underset{k_{off}}{\rightleftharpoons}}  On.
\end{equation}
\noindent Now following Ref.~\onlinecite{Hu2012}, we associate a variable $D$
with the switching process of the promoter, where $D$ takes the value 0 and 1 for the
\textit{off} and the \textit{on} state of the promoter, respectively.
The stochastic Langevin equation associated with $D$ can be written as \cite{Mehta2008}
\begin{equation}
\label{eq6}
\frac{dD}{dt} = k_{on} R_p (1-D) - k_{off} D + \xi_{D}(t) ,
\end{equation}

\noindent
where $\langle \xi_D(t) \rangle = 0$ and
\begin{eqnarray*}
\langle \xi_D (t) \xi_D (t+\tau) \rangle =  2 k_{off} \langle D \rangle \delta (\tau) ,
\end{eqnarray*}

\noindent with steady state value
$\langle D \rangle= \langle R_p \rangle/(\langle R_p \rangle+K_d)$ and
$K_{d}=k_{off}/k_{on}$.
The Langevin equation is simply a noisy version of the deterministic chemical kinetics,
which on the noise averaged level would yield the average value of the variable $D$
for the \textit{on} state of the promoter.
Following Eq.~(\ref{eq6}), we associate a time scale for the promoter
switching kinetics $\tau_{out}$, where $\tau_{out} = (k_{on} \langle R_p \rangle + k_{off})^{-1}$,
a characteristics of noiseless input model.
Now linearizing Eq.~(\ref{eq6}) and performing Fourier transformation of the linearized
equation, one arrives at \cite{Mehta2008}
\begin{eqnarray}
\label{eq7}
\delta \tilde{D} (\omega) =
\frac{\tilde{\xi}_D (\omega)}{i\omega+\tau_{out}^{-1}}
+
\frac{k_{on} (1-\langle D \rangle)}{i\omega+ \tau_{out}^{-1}}
\delta \tilde{R}_p (\omega) .
\end{eqnarray}

\noindent
Note that the first term on the right hand side of Eq.~(\ref{eq7}) arises due to the noiseless
input model (mean field input of $R_p$) and incorporates only the fluctuations in the
promoter switching kinetics, whereas the second term appears via the noisy input
due to the fluctuations in the $R_p$ level.
We now define the total variance associated with $D$ at steady state for the noisy input
model as
\begin{equation}
\label{eq8}
\sigma^2_{D} = \frac{K_d \langle R_p \rangle}{(\langle R_p \rangle+K_d)^2}
+ \frac{K_d k_{off}}{(\langle R_p \rangle+K_d)^3} \sigma^2_{R_p} ,
\end{equation}

\noindent
where $\sigma^2_{D} = (1/2\pi) \int d\omega \langle | \delta \tilde{D} (\omega) |^2 \rangle$.
Here, the first and the second term on the right hand side of Eq.~(\ref{eq8}) arises due to
noiseless input model and fluctuations in the $R_p$ level, respectively.
At this point, it is important to mention that an almost similar expression for the variance
$\sigma^2_{D}$ was obtained by Hu \textit{et al} in their recent work on the role of input
noise in genetic switch (see Eq.~(14) of Ref.~\onlinecite{Hu2012}).
To be explicit, Ref.~\onlinecite{Hu2012} shows that for a noisy input model, the value of
$\langle D \rangle$ itself changes in comparison to the noiseless input model
(constant $R_p$) which eventually changes the variance.
Thus, considering the kinetics of promoter switching as a simple binary process in the
presence of noisy input one arrives at the aforesaid expression of $\sigma^2_D$
which incorporates the essential features of noiseless input model as well as the
fluctuations in the $R_p$ level (via $\delta \tilde{R}_p (\omega)$, see Eq.~(\ref{eq4b})).
This result suggests that the variance due to the noiseless input model gets modified
in the presence of a noisy input and is in agreement with the result shown in
Ref.~\onlinecite{Hu2012}.

To check how the time scale of the noisy input (fluctuations in the $R_p$ level) affects
the promoter switching kinetics, we define noise associated with the promoter
switching at steady state for the noisy input model as $\eta_D = \sigma_D/\langle D \rangle$,
where
\begin{equation}
\label{eq9}
\eta_D = \left [
\frac{K_d}{\langle R_p \rangle}
+ \frac{K_d k_{off}}{(\langle R_p \rangle+K_d)} \eta^2_{R_p}
\right ]^{1/2} .
\end{equation}


\begin{figure}[!t]
\begin{center}
\includegraphics[width=0.75\columnwidth,angle=-90,bb=48 62 590 792]{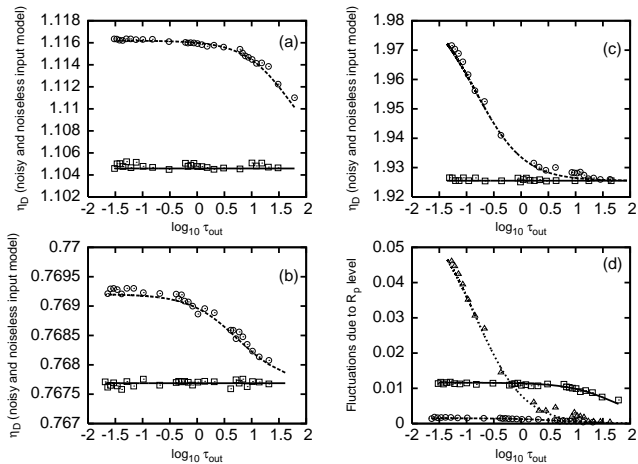}
\end{center}
\caption{(a-c) Plot of $\eta_D$ as a function of promoter switching
time scale $\tau_{out}$, for $\kappa_{11} = 4 \times 10^{-4}, 10^{-2}$ and 0.4,
respectively.
The solid (with open squares) and the dashed (with open circles) lines are for
noiseless and noisy input model, respectively.
(d) Contributions due to fluctuations in the $R_p$ level.
The solid (with open squares), dashed (with open circles) and dotted (with
open triangles) lines are due to $\kappa_{11} = 4 \times 10^{-4}, 10^{-2}$
and 0.4, respectively.
In all the panels, lines are drawn using Eq.~(\ref{eq9}) and the symbols
are generated using Gillespie simulation \cite{Gillespie1976,Gillespie1977}.
}
\label{npms}
\end{figure}

\noindent The first and the second term on the right hand side of Eq.~(\ref{eq9})
is due to the noiseless input model and the fluctuations in the $R_p$ level,
respectively, as suggested by Eq.~(\ref{eq8}).
In Fig.~\ref{npms}, we show the dependence of $\eta_D$ on the promoter switching
time scale $\tau_{out}$ for 1:1 system for low, intermediate and high values of
$\kappa_{11}$. The three values of $\kappa_{11}$ have been chosen from the
monofunctional, cross-over and bifunctional regime, respectively, of the TCS
signal transduction motif (see Fig.~\ref{noise11}a).
Fig.~\ref{npms} suggests that for low and high values of $\kappa_{11}$, the fluctuations
associated with the promoter switching kinetics due to noisy input model are higher
(dashed line with open circles in Figs.~\ref{npms}a,\ref{npms}c) as the fluctuations
due to $R_p$ level ($\eta_{R_p}$, see also Fig.~\ref{noise11}a) at these parameter
regimes are high. However, for the intermediate $\kappa_{11}$ value, the fluctuations are
minimum (dashed line with open circles in Fig.~\ref{npms}b) as the TCS maintains a
minimum noise level at this $\kappa_{11}$ value.
For reference, we show the fluctuations associated with the noiseless input model in
Figs.~\ref{npms}a-\ref{npms}c (solid line with open squares) which clearly
explains enhancement of noise for the noisy input model due to fluctuations in
the $R_p$ level (Fig.~\ref{npms}d).

Fig.~\ref{npms}d shows that as $\tau_{out}$ increases, the contribution due
to the $R_p$ level fluctuations in the promoter switching kinetics decreases, which is a
general trend for all the $\kappa_{11}$ values. In the limit of fast promoter switching rate
(low $\tau_{out}$) compared to the time scale of the $R_p$ level fluctuations ($\tau_{in}$),
$\tau_{out} \ll \tau_{in}$. At this limit, the contribution due to noisy input is high and
the output of the network (TCS) affects the promoter switching kinetics maximally. On the
other hand, when the promoter switching rate is slow (high $\tau_{out}$) compared to the
variation of network output time scale such that $\tau_{out} \gg \tau_{in}$, the network output
exerts a mean field effect on the promoter switching rate. At this limit contribution due to the
noisy input reduces drastically (second term on the right hand side of Eq.~(\ref{eq9})) and
one recovers the behavior of the noiseless input model.


\section{Conclusion}

To conclude, we have provided a generic description for the calculation of noise due to
post-translational modification in the bacterial TCS. From exact analytical calculation within the
purview of linear noise approximation, it is possible to quantify the steady state noise
for the single pair and for the branched pathways. For the single pair system, our analysis
shows the effect of bifunctionality of SK on noise generation and can differentiate the
mono- and the bifunctional domain in the noise profile. The calculation for the branched pathways
shows enhancement and reduction of noise for the composite system in terms of extra
phosphate outflux and influx, respectively. Our analysis suggests that in \textit{one to many}
system as in the chemotaxis pathway of \textit{E. coli}, enhancement of fluctuations happens
due to extra outflux of phosphate group within the members of TCS. On the other hand,
for \textit{many to one} system mimicking the quorum sensing network of \textit{V. cholerae}, an
optimal level of noise can be maintained via extra influx of phosphate group. To maintain
such low noise activity, SK of \textit{V. cholerae} phosphotransfer circuit might prefer to operate
in the cross over domain.

The motif of TCS in the bacterial kingdom reliably transmits the information of the
changes made in the extracellular environment within the cell. This happens via the
formation of the pool of $R_p$ which acts as a transcription factor for several genes
including the genes encoding the TCS. The molecular fluctuations due to the
post-translational modification during the formation of $R_p$ play an important role
in the fluctuations of the gene expression mechanism. While acting as a transcription
factor the noise due to $R_p$ level fluctuations acts as a noisy input in the gene
expression mechanism. On ther hand, the promoter activation/inactivation mechanism is
characterized by the intrinsic molecular fluctuations. Keeping this in mind, we have
investigated the possible role of the network output on the promoter switching kinetics.
The fluctuations associated with the promoter switching mechanism have been quantified
by the total noise at steady state associated with the active state of the promoter. Our
analysis suggests that the total noise $\eta_{D}$ at steady state is comprised of two parts;
the first part arises due to the noiseless input model while the second part is due to the
noisy contribution of the TCS network output. If the fluctuations in the $R_p$ level occur
on a faster time scale, it hardly affects the process of transcription as the DNA promoter
activation/inactivation mechanism gets weakly affected. In such a situation, the transcription
factor $R_p$ exerts a mean field effect in the process of transcription and fluctuations
in the promoter switching kinetics are predominantly governed by the intrinsic molecular
noise, a typical characteristics of the noiseless input model. On the other hand, when the
time scale of $R_p$ fluctuations is slower/comparable than the promoter switching rate,
it exerts a considerable effect on the promoter switching mechanism. In other words, when
the fluctuations in the $R_p$ level maintain an optimal level and is comparable with the
time scale of the DNA promoter switching rate, the latter gets highly affected by the
changes made in the extra-cellular environment which has been reliably transmitted
through the TCS. The formalism we present in this work gives an idea of the optimal level
of fluctuations within the TCS which is necessary for reliable transmission of signal to
control the regulation of biochemical switch present within the bacterial cell.


\begin{acknowledgments}
We express our sincerest gratitude to Indrani Bose for fruitful discussion.
AKM and AB are thankful to UGC (UGC/776/JRF(Sc)) and CSIR (09/015(0375)/2009-EMR-I),
respectively, for research fellowship.
RM acknowledges funding from Academy of Finland (FiDiPro scheme).
SKB acknowledges support from Bose Institute through Institutional Programme
VI - Development of Systems Biology.
\end{acknowledgments}


\end{document}